# MEMPSEP II. - Forecasting the Properties of Solar Energetic Particle Events using a Multivariate Ensemble Approach

Maher A. Dayeh[1,2,*], Subhamoy Chatterjee[3,*], Andrés Muñoz-Jaramillo[3], Kimberly Moreland[2,1,4,5], Hazel M. Bain[4,5], Samuel Hart[2,1]

[1]Southwest Research Institute, San Antonio, TX
[2]The University of Texas at San Antonio, San Antonio, TX
[3]Southwest Research Institute, Boulder, CO
[4]Cooperative Institute for Research in Environmental Sciences, University of Boulder, CO
[5]Space Weather Prediction Center, NOAA, Boulder, CO

Corresponding author: Maher A. Dayeh, maher.aldayeh@swri.edu
*Equal contribution.

**Key Points:**

- An end-to-end deep neural network model ensemble, trained on remote sensing and in-situ datasets, is used to forecast SEP properties.
- Utilizing model ensemble maximizes the utilization of an imbalanced dataset and enhances forecasting confidence with uncertainty estimates.
- SEP occurrence probabilities are utilized as weights in the loss function to enhance regression performance for events.

**Abstract**

Solar Energetic Particles (SEPs) form a critical component of Space Weather. The complex, intertwined dynamics of SEP sources, acceleration, and transport make their forecasting very challenging. Yet, information about SEP arrival and their properties (e.g., peak flux) is crucial for space exploration on many fronts. We have recently introduced a novel probabilistic ensemble model called the Multivariate Ensemble of Models for Probabilistic Forecast of Solar Energetic Particles (MEMPSEP). Its primary aim is to forecast the occurrence and physical properties of SEPs. The occurrence forecasting, thoroughly discussed in a preceding paper (Chatterjee et al., 2023), is complemented by the work presented here, which focuses on forecasting the physical properties of SEPs. The MEMPSEP model relies on an ensemble of Convolutional Neural Networks, which leverage a multi-variate dataset comprising full-disc magnetogram sequences and numerous derived and in-situ data from various sources. Skill scores demonstrate that MEMPSEP exhibits improved predictions on SEP properties for the test set data with SEP occurrence probability above 50%, compared to those with a probability below 50%. Results present a promising approach to address the challenging task of forecasting SEP physical properties, thus improving our forecasting capabilities and advancing our understanding of the dominant parameters and processes that govern SEP production.

# 1. Introduction

Solar energetic particles (SEPs) are an integral component of Space Weather. SEPs in the inner heliosphere comprise different origins, acceleration, and transport mechanisms that alter their properties significantly (see review by Desai and Giacalone (2016)). SEPs are generally classified into two main classes, impulsive and gradual, which correspond to the way SEPs are accelerated. Impulsive SEPs (ISEPs) are generated during explosive flare phases (Reames, 1999), while gradual SEPs (GSEPs) are produced by diffusive acceleration processes at traveling interplanetary (IP) shocks (Cane, 1988), which are associated with the expulsion of coronal mass ejections (CMEs), or develop at corotating IP structures (e.g., Gosling and Pizzo (1999); Mason et al. (2008)). Although both ISEPs and GSEPs exhibit distinct properties when considered individually, the majority of SEP events observed in the heliosphere display a combination of both properties, with large event-to-event variability. This is attributed to the complex environment surrounding the origin (Mason et al. (2004); Desai et al. (2006); Dayeh et al. (2009)), acceleration (e.g., Desai et al. (2016), and propagation (e.g., Mason et al. (2012)) of SEPs, leading to three unresolved primary concerns, with a mounting body of work continuing to understand these major processes (see review by Zhang et al. (2021) and references therein).

The first issue pertains to the origin of SEPs. Observations of exceptionally rare solar wind elements in SEP events provide compelling evidence that CME-driven IP shocks selectively accelerate material from a suprathermal pool consisting of heated solar wind, coronal material, remnants of solar transient events, and more. (see Gloeckler (2003); Desai et al. (2006); Mason et al. (2004); Dayeh et al. (2009, 2017); Alterman et al. (2023); Wijsen et al. (2023)). Quantifying the contributions of these different seed populations to SEPs is crucial for understanding the variability observed from one event to another. The second challenge lies in understanding the acceleration of SEPs. Diffusive shock acceleration (DSA) is considered the primary candidate of SEP acceleration at IP shocks, comprising the shock-drift mechanism at quasi-perpendicular shocks (Decker, 1981), and the first-order Fermi mechanism at quasi-parallel shocks (Lee, 1983); however, observations showed that theory can only explain certain behaviors of SEPs and does not provide a universal explanation for a plethora of mixed cases. The third aspect requiring attention is SEP transport. Scattering within the IP medium plays a significant role in shaping SEP observations at 1 AU. Studies have revealed evidence of heavy-ion scattering by Alfven waves generated by streaming protons accelerated at shocks, indicating that IP scattering is sometimes dominated by a dynamic wave spectrum rather than a universal background spectrum (e.g., (Tylka et al., 2005; Ng et al., 1999)). Work by (Cohen et al., 2005; Mason et al., 2006; Desai et al., 2016) attributed SEP rigidity-dependent scattering as particles propagate through the corona and IP medium, or to substantial cross-field diffusion near the Sun. However, SEP scattering alone does not fully account for the puzzling variability observed in SEP properties at 1 AU.

SEP hazards in the near-Earth environment are of multiple folds. The first and foremost important risk is the radiation hazard for astronauts in Human spaceflight (e.g., Cucinotta (2022) and references therein), and passengers on high-flying polar flight routes (Beck et al., 2005). SEPs are intense ionizing radiation that could have acute or chronic effects on humans. Acute effects cause radiation sickness that puts astronaut lives at immediate risk. Chronic dose effects include carcinogenesis, neural damage, and degenerative tissue damage diseases affecting major organs. In addition to the human element, high-energy penetrating SEPs are known to cause irreversible malfunctions in spacecraft electronics (e.g., Webb and Allen (2004)). While heavy ions cause most damage to electronics, protons account for most radiation hazard because they are abundant and can penetrate easily through spacecraft manuscript submitted to *Space Weather*shielding. To mitigate SEP radiation hazards, space designers rely on models of the extreme space environment to estimate the radiation protection calculations of instrumentation and crew protection. This prediction is often done for several years ahead of the anticipated missions. While this task is feasible for the more-stable and seasonal energetic sources such as the galactic cosmic rays (GCRs) and anomalous cosmic rays (ACRs), it is not applicable to SEPs, not only because of their transient nature, but also due to the large event-to-event variability in SEP properties, such as their peak intensities, spectral indices, elemental abundances, maximum energy attained by particles, etc. Forecasting SEPs (e.g., Dayeh et al. (2010); Kozarev et al. (2010)) and their properties, such as peak flux, event duration, composition, etc., ahead of their arrival is thus critically informative to mitigate SEP radiation hazards. With the emergence of machine learning algorithms in space weather applications, physical processes are often ignored at the expense of trend detection and correlation analysis. While machine learning (ML) shows promising results in predicting space weather events, a complete understanding of the underlying physics remains challenging. Therefore, it is essential to recognize the importance of utilizing both ML and traditional physics-based approaches to gain comprehensive insights into space weather phenomena. Relating SEP characteristics to their precursors has been the subject of substantial research (e.g., Gopalswamy et al. (2004); Lara et al. (2003); Garcia (2004b, 2004a); Belov et al. (2005); Richardson et al. (2018); Anastasiadis et al. (2017); Kahler and Ling (2018); Papaioannou et al. (2018); Dayeh et al. (2018); Lario et al. (2023)). This research revealed correlative relationships that have enabled the development of qualitative predictive models of SEP phenomena. For instance, (Dayeh et al., 2018) explored the parameter space of ESPs with their shock properties and found that shock speed is directly related to the associated ESP peak, a result that is also reached by (Lario et al., 2023) for a different set of events. (Richardson et al., 2018) leveraged CME speeds and magnetic connectivity to estimate the SEP peak proton intensity. (Posner, 2007) utilized relativistic electrons to forecast SEP arrival times and profiles. However, to enhance these predictive models, there is a pressing need for a comprehensive multivariate analysis that considers all available relevant parameters and interwoven physical processes, including the conditions of solar eruptions, as also detailed in (Bain et al., 2021). This should also advance the capability to have probabilistic forecast models as opposed to simple binary-outcome models. For a detailed review of current models, see (Whitman et al., 2022).

Ensemble models enable proper estimation of prediction uncertainty and ultimately improves correct decisions in critical situations. A new model, the Multivariate Ensemble of Models for Probabilistic Forecast of Solar Energetic Particles" or MEMPSEP, has recently been developed and extensively validated (Chatterjee et al., 2023). The work also investigated the advantage of utilizing an ensemble of models to produce confident and reliable outcome. MEMPSEP enables forecasting the occurrence of an SEP event and in the case of event forecast, it provides the forecast of its physical properties such as peak intensity, duration, and elemental composition, among other user-defined properties. This work complements (Chatterjee et al., 2023) and describes the regression branch of MEMPSEP which allows SEP properties forecasting. Section 2 briefly describes the data set, sections 3, 4, and 5 describe the ensemble of models setup, section 6 presents the results, and section 7 if for discussion and conclusions.

Table 1.  Energy ranges for SEP Peak flux

| Peak 1 | Peak 2 | Peak 3 | Peak 4 | Peak 5 |
|---|---|---|---|---|
| $\geq$5 MeV | $\geq$10 MeV | $\geq$30 MeV | $\geq$60 MeV | $\geq$100 MeV |

## 2 Dataset

The dataset used in MEMPSEP development is extensively described in (Moreland et al., 2023). The data set is comprised of a large set of parameters from remote (e.g., images) and in situ (e.g., time series and pre-eruptive event properties) sources. In this context, ingested data are the "predictors" and the outcome "targets" are the SEP parameters. The sections below provide a brief description of both.

### 2.1 Predictors

It has been already established in Paper-I by Chatterjee et al. (2023) that both in-situ and remote-sensing parameters are crucial in reliably forecasting the occurrence of SEPs. We use a sequence of full-disk line-of-sight magnetograms (SoHO/MDI + SDO/HMI) prior to flare onset as remote-sensing input. For the sequence, we use a sampling interval of 6 hours over a period of 3 days and rebin full-disc magnetograms to a size of 256*pixels* ×256*pixels* creating a data cube of shape 256 ×256 ×13. We use a set of in-situ properties namely: solar wind (SW) temperature, SW velocity, SW density, interplanetary magnetic field (IMF) field strength (B), IMF Bx, By and Bz components, as well as particle fluxes and ratios (Fe/O, H, O, and Fe) calculated prior to every flare onset. We emphasize here that the data is pre-event and the forecasting is thus a true forecast as opposed to nowcasting. Additionally, we use Wind/Waves radio-burst time-frequency images (432 [time]×80 [frequency]), L1-electrons (7-channels of 8640 time points), and X-ray (2-channels of 1441 time points) time series as in-situ inputs. For details about the data preparation see Paper-III by Moreland et al. (2023).

## 2.2 Targets

We set forecast targets as 7 SEP properties, namely, integrated peak flux (in 5 energy bands determined from GOES; see Table 1), onset time of SEP w.r.t. onset of flare (named as 'FlrtoSEP dt') and duration of SEPs. It must be noted that the onset time and duration of SEPs named as "evens" are available only when the peak flux (>10 MeV) exceeds a threshold of 5 p.f.u. within 6 hours from the flare onset. For non-events, we set those parameters to a constant value of 720 hours (see definition of 'events' and 'non-events' in Paper-I by Chatterjee et al. (2023)). Before fitting a model we take the logarithm of targets to minimize the effect of outliers.

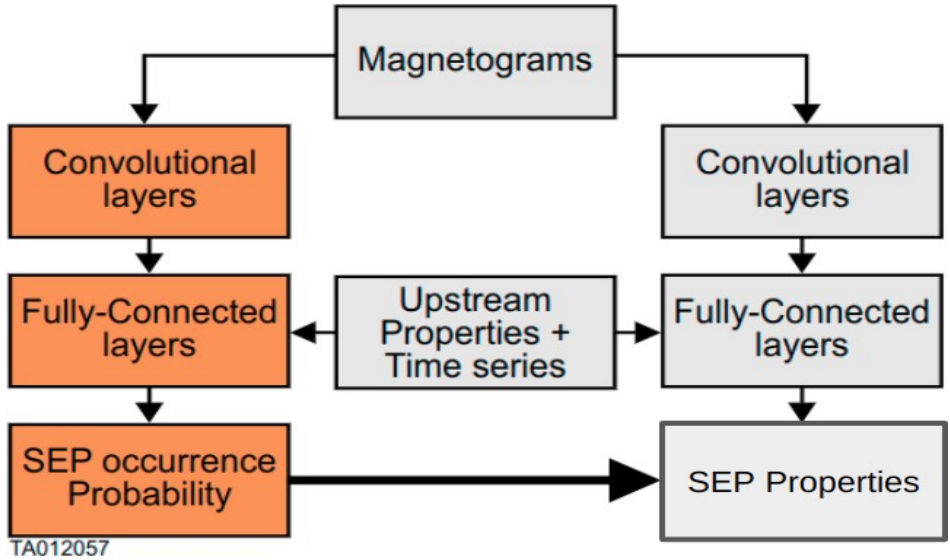

**Figure 1.** CNN model to perform SEP property regression using feedback from the pre-trained classification model. We first train the classification branch (for detailed architecture see Paper-I by Chatterjee et al. (2023)) depicted in orange and then freeze weights and biases of that while training the regression branch shown in gray. The outcome of the classification branch (i.e. the SEP occurrence probability)

penalizes the loss function of the regression branch depicted by the horizontal arrow at the bottom of the figure.

## 3 Model description

### 3.1 Architecture

MEMPSEP workhorse is an ensemble of Convolutional Neural Network (CNN; see LeCun et al. (2015)) architecture that ingests multiple inputs such as images, scalar parameters and time series (Figure 1) and forecasts the 7 SEP properties. The CNN processes the input images image sequences with repeated 2D convolution, batch-normalization, non-linear activation and max-pooling. The flattened layers are concatenated with scalar parameters and features extracted from time series data, and passed through dense (fully connected) layers to produce the SEP properties. Dropout is used before the last layer causing random disconnections between nodes and helping in regularization of the model. It must be noted that the temporal information of the time series data is fused in the initial layers of the CNN and then broken into multiple feature layers that further passed onto subsequent layers. The model architecture is similar to that of the model for forecasting SEP occurrence (see Paper-I by Chatterjee et al. (2023)) except for the lack of any non-linear activation at the end. We avoid any non-linearity after the final layer as activation functions like 'ReLU' may cause a dying-ReLU problem and drive the model optimization toward all-zero predictions.

### 3.2 Optimization

To optimize the model's weights and biases we minimize the mean squared error (MSE) loss described by:

$$MSE = \frac{1}{NB}\sum_i \sum_{j=1}^{B}(target_{i,j} - predicted_{i,j})^2 \qquad (1)$$

where i ∈ {peak1, peak2, peak3, peak4, peak5, onset, duration}, N stands for the number of targets i.e. 7 and j runs over a batch of size B. We use a training-validation split of 4:1 (1020 (non-events) + 680 (events) for training and 255 (non-events) + 170 (events) for validation) and stop training the model when the validation accuracy stops improving (Figure 2). We call this model

'non-gated'. We also make use of SEP occurrence probability produced by another CNN model depicted in Paper-I by Chatterjee et al. (2023). At this stage, we freeze the weights and biases that were learned during the optimization of the SEP classification branch and just train the regression branch penalizing the loss function of the classification outcome through a custom probability-weighted mean-squared error (PSE) loss described as follows:

$$PSE = \frac{1}{NB}\sum_i \sum_{j=1}^{B} p_j (target_{i,j} - predicted_{i,j})^2 \qquad (2)$$

$p_j$ represents the predicted SEP occurrence probability by the frozen classification branch for training instance $j$. We adapt this approach so that the model is driven to match the target better for the instances that are predicted to have a higher chance of SEP occurrence. We call this model 'gated'.

## 4 Ensemble of Models

The fact that there is a significant class imbalance between SEP events and non-events is widely recognized as a challenging issue for the effectiveness of Machine Learning-based SEP prediction models (Kasapis et al., 2022). To address this problem, we adopt a strategy where we create several training and validation sets by combining substantially diverse non-events with the same set of events. To achieve this, we begin by assessing the distribution of flare peak strength in events using an equal-frequency histogram with 10 bins. Next, we utilize the edges of these bins to randomly select non-events, aiming for a number that is 1.5 times greater than the number of events in each bin. This selection process is repeated ten times, ensuring minimal overlap between the chosen non-events for each bin. We have approximately 850 event samples (85 per bin), and for each draw, we randomly select around 1275 non-events (approximately 127 per bin). So, in total, for 10 draws, we end up with close to 10,000 non-event samples. These non-event samples, along with the 850 events, are used to train and validate a model (Figure 2). By doing this, we create an ensemble of 10 models. This approach is necessary because training a single model would have resulted in a severe class imbalance of 1 event to 15 non-events.

## 5 Design of Test set

We design a test set for the forecast of SEP occurrence probability (also used in Paper-I by Chatterjee et al. (2023)) that represents all the flare classes and is not affected by the solar cycle phase in event selection (Figure 3). We select 5 events and 10 non-events perflare bins for all the 10 bins and also ensure that a year is dropped from random draw When it appears more than times. We also ensure that the test set events are well separated in time from training and validation sets. This test set allows us to provide an unbiased estimate of model-ensemble performance.

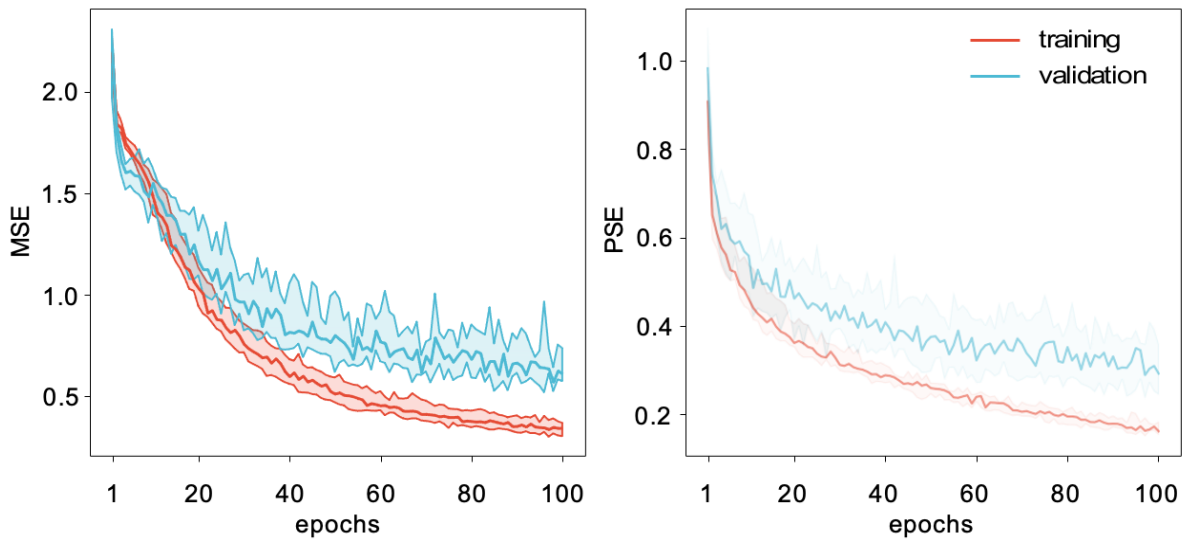

**Figure 2.** Covergence of loss function with epoch. Left panel depicts MSE loss for training and validation set as a function epoch. Right panel shows PSE loss for training and validation set as a function of epoch. The shaded regions depict the uncertainty (10th - 90th percentile produced by model ensemble.)

## 6 Regression Results

For every data point on the test set, we acquire the model-ensemble inference for both 'non-gated' and 'gated' regression. We produce scatter-plots of target vs. predicted values of SEP peaks (1–5), FlrtoSEP dt, and SEP duration and color the points according to the median SEP occurrence probability ($p$) predicted by the model ensemble. We also add different symbols to distinguish between the points representing SEP 'events' and 'non-events' ground truths (Figures 4,5).

We use different metrics to quantify the regression performance, namely, Pearson Correlation Coefficient (CC), Root Mean Squared Error (RMSE), and R2 score for groups of points having $p \geq 0.5$ and $p < 0.5$ R2 score, also known as the coefficient of determination and represents the fraction of the variability in the regression target that can be explained by the model. Thus an R2 score close to 1 is desirable. An R2 score less than or equal to zero is an indicator that the model is not performing any better than just an average of the regression target.

We observe a general tendency of the points to get closer to the 'y=x' line for higher occurrence probabilities. Except for 'FlrtoSEP dt' and 'SEP duration' we find higher contrast between the metric corresponding to those two groups ($p \geq 0.5$ and $p < 0.5$) for 'gated' regression (Table 3) as compared to 'non-gated' (Table 2). For both the regression types, we find the CC between the logarithm of the target and the predicted ensemble-median of SEP peaks to be statistically significant and $> 0.8$. Also, RMSE ranges between 0.44–0.68 signifying a factor of 3–5 difference between target and predicted values. For 'non-gated' and 'gated' regression, we find a minimum R2-score of 0.52 and 0.63 respectively for SEP peaks with $p \geq 0.5$. For 'FlrtoSEP dt' and 'SEP duration' we generally find worse performance than other regression targets and also observe that the 'non-gated' models perform slightly superior to 'gated' ones.

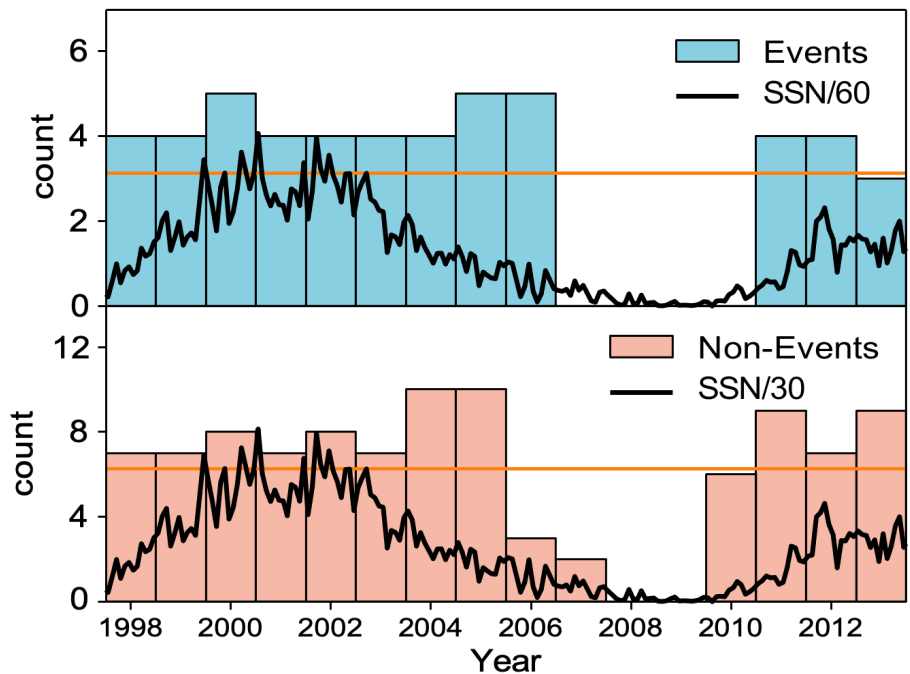

**Figure 3.** The top panel of the yearly histogram displays events in sky-blue, while the bottom panel represents non-events shown in pink. The horizontal orange line indicates the uniform event count per bin if the events were evenly distributed. The black curve depicts the smoothed monthly sunspot number cycle. From the graph, we can observe that event counts do not follow the modulation of the sunspot number cycle, ensuring an unbiased estimation of model performance on the test set.

To understand the ability of the model ensemble to identify 'events' and 'non-events' as separate populations, we visualize the probability density of the predicted values and check if there is a significant difference between those for 'non-events' and 'events'. We perform an independent one-tailed t-test between those and find a significant difference for all the targets (p-values<0.05; see Table 4) with 'non-gated' and 'gated' models. Lastly, to examine the effectiveness of the 'gated' model, we plot skill scores as functions of probability threshold for all regression targets (Figure 6,7). We find the 'gated' model to cause improvements for most of the SEP peaks as compared to 'non-gated' for higher values of the threshold with a crossover happening at around a probability of 0.4. For 'FlrtoSEP dt' and 'SEP duration' we find that crossover happening for RMSE and R2 scores with 'non-gated' model having slightly superior scores as compared to 'gated'. Although it should be noted R2 score for those two targets is close to 0 or negative indicating the 'gated' and 'non-gated' model performance to be no better that a constant outcome of average target value.

## 7 Discussion and Conclusions

It can be observed that skills scores for 'FlrtoSEP dt' and 'SEP duration' are significantly worse than those for other regression targets. False positives depicted by triangles that are dark ($p \geq 0.5$) in the scatterplots play a major role in worsening the Skill scores for 'FlrtoSEP dt' and 'SEP duration'. Even though the model ensemble is able to predict SEP Peak values for those non-events, there may not be enough information in the inputs to predict such large values of 'FlrtoSEP dt' and 'SEP duration'. This could also be the under-represented set of such non-events in the training set that cross the 5 p.f.u. cutoff for SEP flux (> 10 MeV) but takes longer than 6 hours. Thus the model ensemble tends to wrongly predict shorter 'FlrtoSEP dt' and 'SEP duration' for those. The 'gated' model producing even smaller values for those targets corresponding to false positives worsens the skill scores further as compared to 'non-gated'. Even though we evaluate uncertainties in prediction through the model ensemble, it is unknown when those uncertainties converge. Further investigations are necessary to ensure well-calibrated uncertainties.

In summary, we make use of the class imbalance between SEP events vs. non-events and train an ensemble of CNNs to forecast a set of SEP parameters. The skills scores show that the ensemble is able to make tighter predictions on SEP properties for the test set data points with SEP occurrence probability ≥0.5 as compared to those with probability < 0.5. We experiment by applying weightage to the loss function with occurrence probability and observe improvement as increased contrast between the metrics for those two probability groups.

With our current dataset and approach, the regression results are not as good for 'FlrtoSEP dt' and 'SEP duration' as for the remaining properties. This problem should be minimized as we acquire more data across the range for each regression target. Also, having the model-ensemble members trained on a balanced dataset, we plan to perform interpretability analysis in the future to decrypt the inner working of the models. Additionally, we would like to make further refinements in the generation of model-ensemble in the future and find the convergence point for the estimated uncertainties.

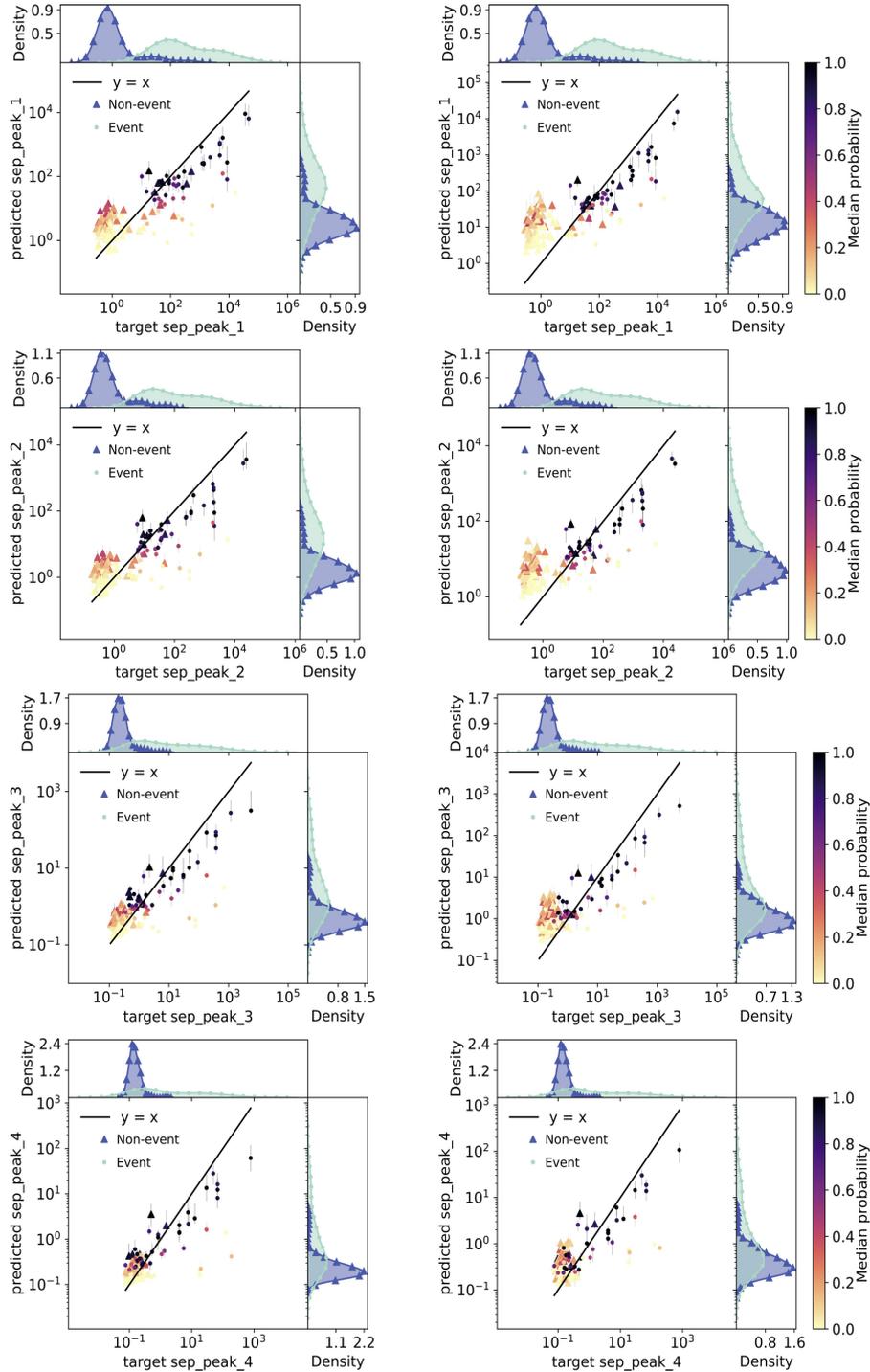

**Figure 4.** Regression of SEP peak flux in different energy bands. Outcome of non-gates and gated models are depicted through left and right columns respectively. Rows represent different SEP peaks. Each scatter plot depicts the target vs median of predicted values with Model-ensemble on the test set. The points are colored by the predicted SEP probability through the classification branch. Ground-truth of event class is depicted in different symbols (triangular for non-events; circular for events). The error bars represent th–e92–5th-75th

percentile of predictions by the model- ensemble. Marginals represent the probability density of the target and predicted medians for events and non-events.

**Table 2.** Skill scores for predicted events and non-events (non-gated)

|  | Peak 1 | | Peak 1 | | Peak 1 | | Peak 1 | | Peak 1 | | Time to SEP | | SEP duration | |
|---|---|---|---|---|---|---|---|---|---|---|---|---|---|---|
|  | p≥0.5 | p<0.5 | p≥0.5 | p<0.5 | p≥0.5 | p<0.5 | p≥0.5 | p<0.5 | p≥0.5 | p<0.5 | p≥0.5 | p<0.5 | p≥0.5 | p<0.5 |
| CC | 0.83 | 0.48 | 0.88 | 0.48 | 0.92 | 0.48 | 0.92 | 0.51 | 0.9 | 0.5 | 0.31 | 0.35 | 0.09 | 0.22 |
| RMSE | 0.68 | 0.92 | 0.61 | 0.84 | 0.53 | 0.67 | 0.48 | 0.55 | 0.46 | 0.51 | 1.44 | 1.22 | 0.82 | 0.63 |
| R2 score | 0.52 | 0.2 | 0.64 | 0.2 | 0.76 | 0.21 | 0.77 | 0.22 | 0.74 | 0.2 | 0.04 | 0.1 | -0.02 | 0.0 |

**Table 3.** Skill scores for predicted events and non-events (gated)

|  | Peak 1 | | Peak 1 | | Peak 1 | | Peak 1 | | Peak 1 | | Time to SEP | | SEP duration | |
|---|---|---|---|---|---|---|---|---|---|---|---|---|---|---|
|  | p≥0.5 | p<0.5 | p≥0.5 | p<0.5 | p≥0.5 | p<0.5 | p≥0.5 | p<0.5 | p≥0.5 | p<0.5 | p≥0.5 | p<0.5 | p≥0.5 | p<0.5 |
| CC | 0.86 | 0.36 | 0.91 | 0.38 | 0.94 | 0.39 | 0.92 | 0.42 | 0.89 | 0.39 | 0.3 | 0.28 | 0.24 | 0.2 |
| RMSE | 0.6 | 1.22 | 0.53 | 1.1 | 0.48 | 0.81 | 0.46 | 0.62 | 0.44 | 0.57 | 1.41 | 1.75 | 0.81 | 0.94 |
| R2 score | 0.63 | -0.41 | 0.73 | -0.34 | 0.81 | -0.14 | 0.79 | 0.02 | 0.75 | 0.02 | 0.07 | -0.85 | 0.01 | -1.27 |

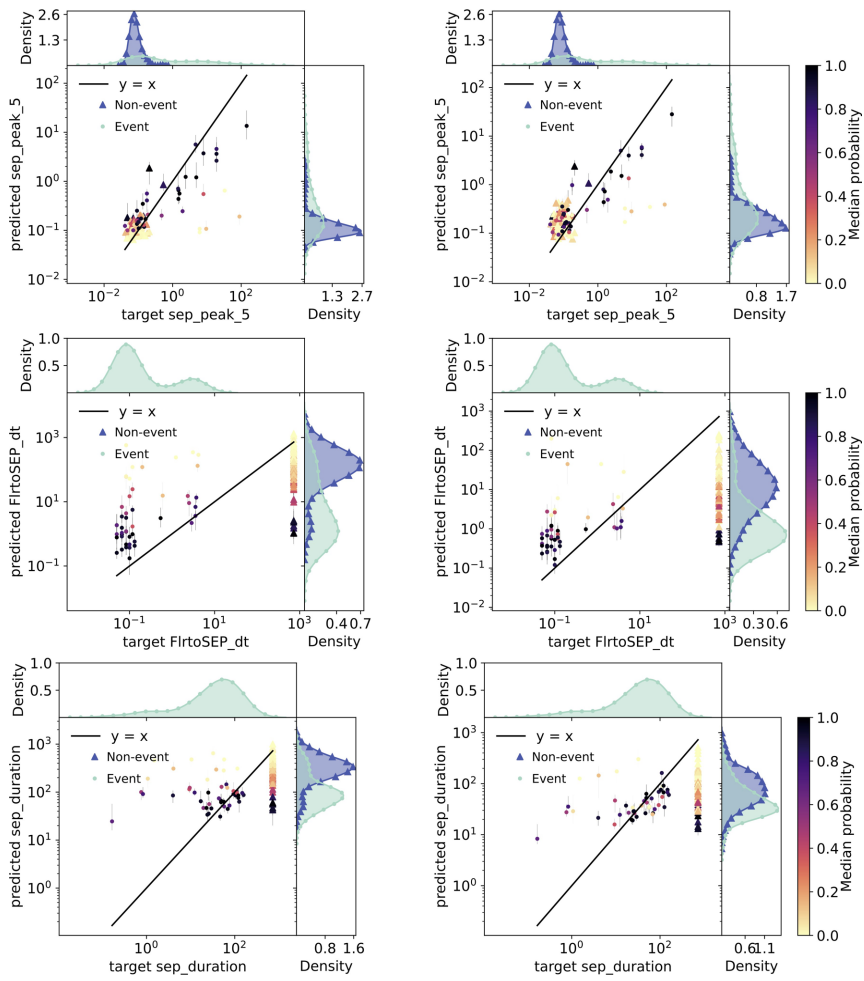

**Figure 5.** Same as Figure 4 except that here the Regression results are shown for SEP peak flux in the highest energy range, SEP onset time, and duration.

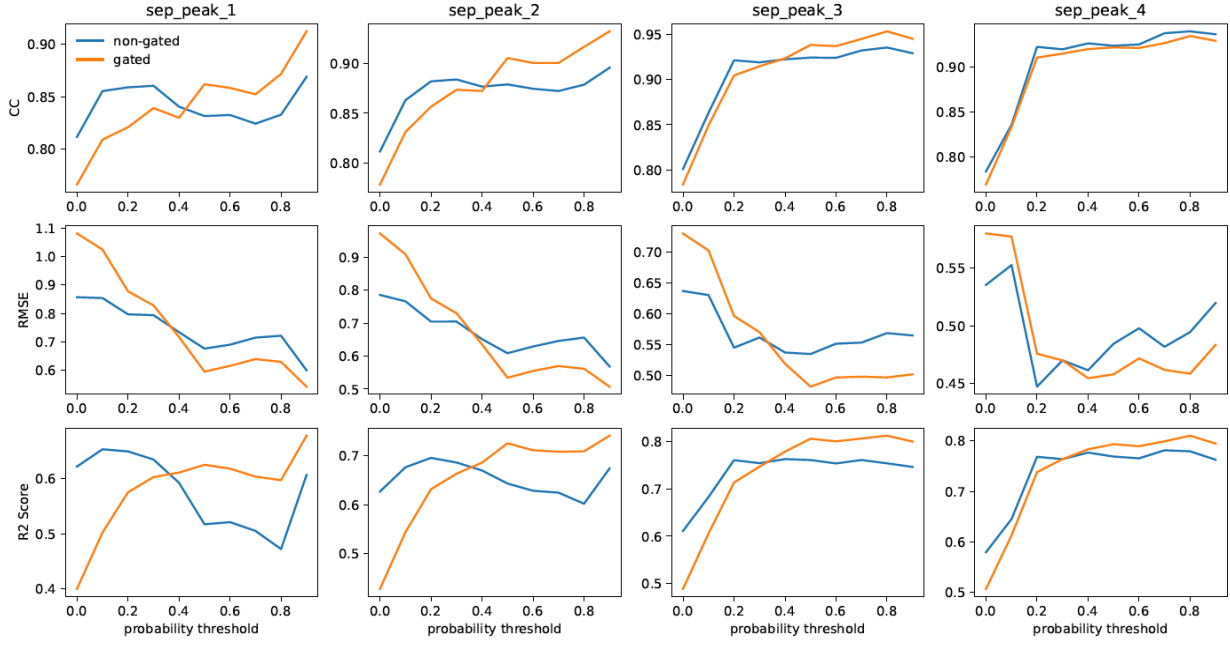

**Figure 6.** Skill scores vs. probability threshold for non-gated and gated regression. The top row depicts a variation of the Pearson correlation coefficient between predicted and target SEP properties as a function of the probability threshold to filter out points with higher predicted SEP occurrence probability. The Middle and bottom rows show variations of Root Mean squared Error (RMSE) and R2 score respectively as a function of the probability threshold. Columns depict different SEP properties. The outcomes of 'Non-gated' and 'Gated' regression models are shown with blue and orange curves respectively.

**Table 4.** p-values for independent t-test between predicted values of events and non-events.

|  | Peak 1 | Peak 1 | Peak 1 | Peak 1 | Peak 1 | Time to SEP | SEP duration |
|---|---|---|---|---|---|---|---|
| Non-gated | 0.021 | 0.023 | 0.013 | 0.014 | 0.008 | 0.0 | 0.0 |
| Gated | 0.033 | 0.023 | 0.019 | 0.027 | 0.027 | 0.001 | 0.0 |

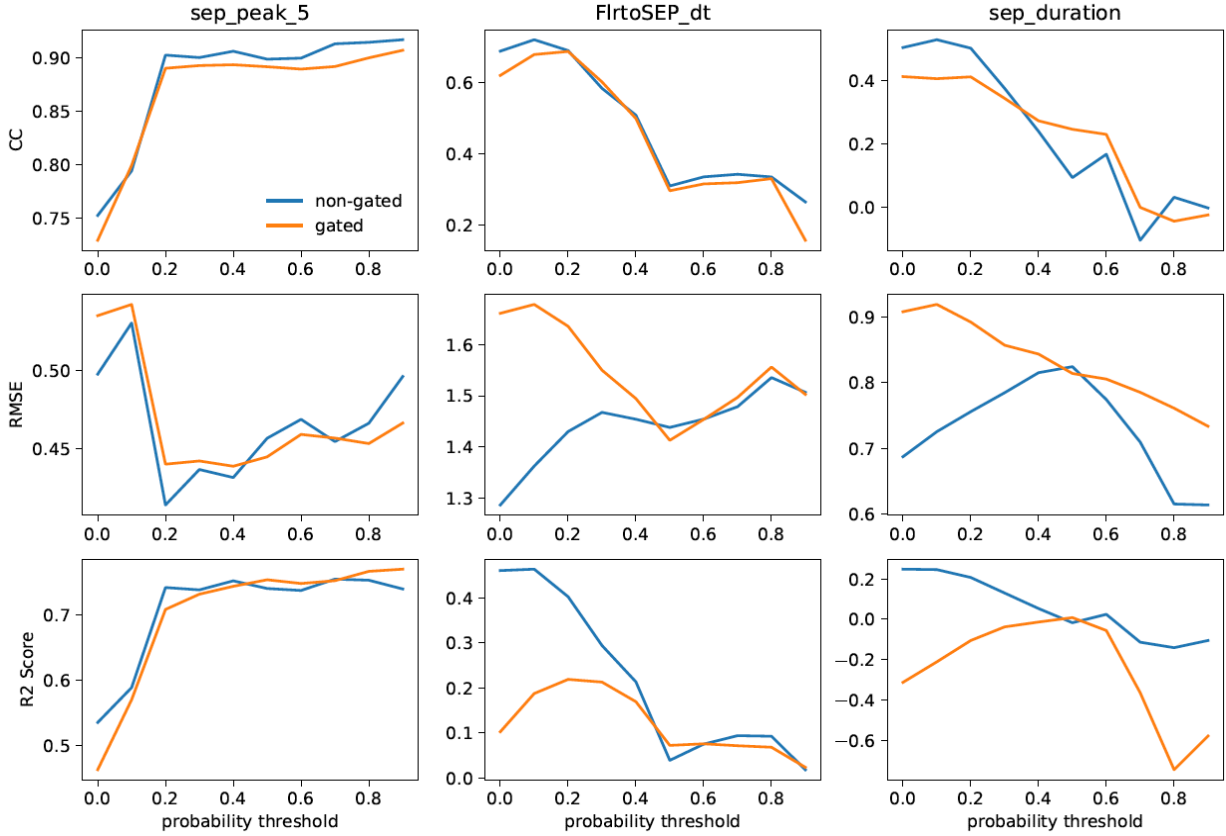

**Figure 7.** Same as Figure 6 except that here the Regression results are shown for SEP peak flux in highest energy range, SEP onset time, and duration.

## 8 Open Research

The dataset used in this paper comes from public repositories (e.g. Heliospheric Event Knowledgebase (HEK), JSOC) and are rigorously described in Paper III by (Moreland et al., 2023). After the publication of this paper, we will make the AI/ML-ready data publicly available through Zenodo. The codebase developed for this work can be accessed from https://github.com/subhamoysgit/MEMPSEP.

## Acknowledgments

This work is mainly supported by SWO2R grant 80NSSC20K0290. Partial support for MAD, KM, and SH comes from SWO2R 80NSSC21K0027, NASA LWS grants 80NSSC19K0079, 80NSSC21K1307, and 80NSSC20K1815.